\documentclass[pre,amsmath,amssymb,twocolumn,superscriptaddress]{revtex4-2}
\usepackage{amsmath,amssymb}
\usepackage[usenames]{color}
\usepackage{amssymb}
\usepackage{grffile}
\usepackage{graphicx}
\usepackage{amsmath, amstext, amssymb, amsfonts, amsxtra}
\usepackage{textcomp}
\usepackage{xspace}
\usepackage{bbm}
\usepackage{bm}
\newcommand{\be}{\begin{equation}}
\newcommand{\ee}{\end{equation}}

\newcommand{\como}{Center for Nonlinear and Complex Systems, Dipartimento
di Scienza e Alta Tecnologia, Universit\`a degli Studi dell'Insubria,
via Valleggio 11, 22100 Como, Italy}
\newcommand{\infn}{Istituto Nazionale di Fisica Nucleare, Sezione di Milano,
via Celoria 16, 20133 Milano, Italy}
\newcommand{\palermo}{Dipartimento di Fisica e Chimica - Emilio Segr\'e, Universit\`a degli Studi di Palermo, via Archirafi 36, I-90123 Palermo, Italy}
\newcommand{\NEST}{NEST, Istituto Nanoscienze-CNR,
Piazza S. Silvestro 12, 56127 Pisa, Italy}

\begin{document}
\date{\today}

\title{Extracting work from random collisions: A model of a quantum heat engine}

\author{Vahid Shaghaghi}
\affiliation{\como}
\affiliation{\infn}
\author{G. Massimo Palma}
\affiliation{\palermo}
\affiliation{\NEST}
\author{Giuliano Benenti}
\affiliation{\como}
\affiliation{\infn}
\affiliation{\NEST}

\begin{abstract}
We study the statistical distribution of the ergotropy and of the efficiency of a single-qubit battery ad of a single-qubit Otto engine, respectively fuelled by random collisions. The single qubit, our working fluid, is assumed to exchange energy with two reservoirs, a non-equilibrium "hot" reservoir and a zero temperature cold reservoir. The interactions between the qubit and the reservoirs is described in terms of a collision model of open system dynamics. The qubit  interacts with the non-equilibrium reservoir (a large ensemble of qudits all prepared in the same pure state) via random unitary collisions and with the cold reservoir (a large ensemble of qubits in their ground state) via a partial swap. Due to the random nature of the interaction with the hot reservoir, fluctuations in ergotropy, heat, and work are present, shrinking with the size of the qudits in the hot reservoir. 
While the mean, ``macroscopic'' efficiency of the Otto engine is the same as in the case in which the hot reservoir is a thermal one,
the distribution of efficiencies does not support finite moments, so that the mean of efficiencies does not coincide with the 
macroscopic efficiency.

\end{abstract}

\maketitle

\section{Introduction}

The development of quantum technologies~\cite{qcbook} is pushing the realm of thermodynamics 
to nanoscale heat engines~\cite{Kosloff2012,Gelbwaser2015,Vinjanampathy2015,Sothmann2015,Goold2016,Benenti2017}.
Here the working fluid can be a particle with discrete energy levels, or in the extreme case even a single qubit.
Such setups challenge the validity of thermodynamic concepts  and raise fundamental questions related to 
the discretness of energy levels, the relevance of quantum coherences, the (possibly strong) coupling 
to non-equilibrium reservoirs~\cite{Scully2003,Huang2012,Abah2014,Rossnagel2014,Hardal2015,Niedenzu2016,Manzano2016,Klaers2017,Agarwalla2017,Niedenzu2018,Cherubim2019,Wang2019,Latune2021},
and the same definition of heat and work, to name but a few.

Collision models~\cite{Rau1963,Alicki1987,Scarani2002,Ziman2002,Ziman2005,Benenti2007,Giovannetti2012,Lorenzo2015,Strasberg2017,DeChiara2018,Pezzutto2019}
(see~\cite{Campbell2021} for a succint entry point to this 
framework and~\cite{Ciccarello2021} for an extensive review)
are an important tool for quantum thermodynamics, as they can be used to conveniently model 
the interaction with reservoirs as unitary tranformations, even in the regime of strong system-reservoir coupling.
It is then possible to address fundamental questions like the relaxation to equilibrium,
the link between information and thermodynamics,
the efficiency of thermodynamic cycles in multi-level engines~\cite{Uzdin2014}, 
and non-Markovian effects.

In this work, we consider a particular kind of non-equilibrium, ``hot'' reservoir, whose interaction with the working fluid 
is modeled by random collisions~\cite{Pineda2007,Pineda2007b,Akhalwaya2007,Gennaro2008,Gennaro2009}. 
Such possibility is quite appealing since by definition
random collisions are a ``cheap'' resource, meaning that no control of type and duration of  system-environment interaction is 
needed. Here the question is whether random collisions can be exploited as a useful resource to perform quantum thermodynamic tasks.

More specifically,  we consider a working medium consisting of a single qubit, alternating collisions with a hot, non-equilibrium reservoir
and a cold, thermal reservoir. The collisions with the hot reservoir, consisting of qudits, are modeled as random unitaries,
while the collisions with the qubits of the cold reservoir are, as usual in the literature, modeled by partial swap operations.
In this protocol, the qubit acts as a 
quantum battery~\cite{Alicki} (see~\cite{Campaioli,Bhattacharje} for reviews), which is charged (discharged) via collisions with the hot (cold) reservoir. 
We fully characterize the process by looking at the statistical distribution of the ergotropies after each collision.

The non-equilibrium reservoir enhances coherences in the qubit system and we analyze their inpact in a quantum Otto cycle.
We show that the macroscopic mean efficiency of this quantum engine after many cycles is the same as for a standard quantum Otto 
cycle (i.e, with a high-temperature reservoir rather than the non-equilibrium one). On the other hand, fluctuations in work and 
heat depend on the dimensionality of the qudits in the non-equilibrium reservoir. Moreover, the distribution of single-cycle 
efficiencies is size-invariant. We derive the efficiency distribution and show that it does not support finite moments, so that 
the mean of efficiencies does not coincide with the macroscopic efficiency.

\section{Ergotropy flow}

\subsection{Charge and discharge of the single-qubit battery}
Let us first analyse the process of charging  and discharging of a single qubit battery interacting with two reservoirs. 
The qubit is charged via an exchange of energy - and coherence -  with a "hot", non-equilibrium, reservoir consisting of a large number of qudits 
(each belonging to a Hilbert space of dimension $\mu$), all identically prepared in a pure state described by the density operator $\hat{\chi}$. 
The qubit interacts with such reservoir via a sequence of pairwise random collisions  with the individual qudit of the  environment.
Such collisions are modelled as a random unitary $\mathcal{\hat R}(L)$ ($L=2\mu$ is the dimension of the joint qubit-qudit  Hilbert space),
drawn from the invariant Haar measure on the unitary group $U(L)$ and conveniently parametrized in terms of the Hurwitz 
representation~\cite{Pozniak1998}.
It is important to note that such collisions are not `weak'', i.e. they can strongly change both the energy and the coherences of the single qubit battery. 

The  battery then dumps its energy into a cold reservoir, consisting of a large number of qubits, all identically prepared in a thermal state $\hat{\vartheta}$. Again the system-environment interaction takes place via a sequence of pairwise collisions between the battery qubit and the environment qubits. 
We assume  each collision to be described by a (unitary) partial swap operation:
\begin{align}
\hat{\mathcal P}(\alpha)=\cos{\alpha}{\hat I}+i\sin{\alpha}\hat{\mathcal S},\quad
\left(0\le\alpha\le\frac{\pi}{2}\right),
\end{align}
where $\hat{\mathcal S}$ is the swap operator:
$\hat{\mathcal S}(|\phi\rangle\otimes|\psi\rangle)\equiv |\psi\rangle\otimes|\phi\rangle$.
If the system state before a collision with the cold reservoir is 
${\hat{\varrho}}$, then after the collision it is 
\begin{align}
{\hat{\varrho}^{\,\prime}=\cos^2\!\alpha\,\hat{\varrho}}+\sin^2\! \alpha\,\hat{\vartheta}+
i\sin{\alpha}\cos{\alpha}[\hat{\vartheta},\hat{\varrho}],
\end{align}
where we have traced over the environment degrees of freedom.
Note that, in the case of complete swap, i.e. $\alpha=\frac{\pi}{2}$, the qubit state $\hat{\varrho}^{\,\prime}$
after the collision is a Gibbs state, which is, as we will explain shortly,  passive. In that case, battery discharging is complete.

Since the ``hot'' reservoir is a non-thermal environment consisting  of pure states interacting with the battery via random unitaries, as we mentioned already the system-environment collisions will modify the battery coherences. In this scenario a convenient quantity to analyse the flux of energy from the hot to the cold reservoir is the battery ergotropy~\cite{Allahverdyan2004}, i.e. the maximum amount of work that can be extracted from the battery via a suitable unitary evolution $\hat{\mathcal U}$ (in our case a single qubit non-dissipative evolution). 

For a system described by a density operator ${\hat\varrho}$ and Hamiltonian $\hat{H}$, the ergotropy is defined as 
\begin{equation}
\mathcal{E}({\hat\varrho},\hat{H})={\rm Tr}({\hat\varrho} \hat{H})- {\rm min} [{\rm Tr}(\hat{\mathcal U}{\hat\varrho}\, \hat{\mathcal U}^\dag \hat{H})],
\end{equation}
where the minimum is taken over all possible unitary transformations $\hat{\mathcal U}$. Given the state $\hat\varrho =\sum_n r_n |r_n\rangle\langle r_n|$ and the Hamiltonian 
$\hat{H}=\sum_n \varepsilon_n|\varepsilon_n\rangle\langle\varepsilon_n|$,
with $r_0\ge r_1\ge...$, and $\varepsilon_0\le \varepsilon_1\le...$, there is a unique state 
\[\hat{\pi}=\hat{\mathcal U}{\hat\varrho}\, \hat{\mathcal U}^\dag =\sum_n r_n |\epsilon_n\rangle\langle\epsilon_n|\]
which minimizes ${\rm Tr}(\hat{\mathcal U}{\hat\varrho}\, \hat{\mathcal U}^\dag \hat{H})$. 
The state $\hat{\pi}$ is called passive, since it cannot deliver any work via the above unitary dynamics.

Given the Bloch spere representation of the qubit state, 
$\hat{\varrho}=\frac{1}{2}\,(\hat{I}+{\bf r}\cdot \hat{\bm{\sigma}})$,
with ${\bf r}=(x,y,z)$ Bloch vector and 
$\hat{\bm{\sigma}}=(\hat{\sigma}_x,\hat{\sigma}_y,\hat{\sigma}_z$) the vector of Pauli matrices,
and the Hamiltonian $\hat{H}=\frac{1}{2}\Delta\hat{\sigma}_z$,
we have $\mathcal{E}=\Delta(r+z)$. The qubit acts as a quantum battery, which can be charged or discharged
via unitary interactions (collisions) with qubits (or qudits) of the environment.

We consider a sequence of charging/discharging cycles.
The reservoirs are assumed to be so large that the system never collides twice with the same environment qudit (qubit).
We can have a pictorial view of the model 
by considering a single qubit colliding in 
sequence with the individual qudits (qubits) of two long chains, corresponding to the hot and cold reservoirs, respectively. 
If $\hat{\varrho}_n$ denotes the system's density operator after $n$ cycles (collisions with each reservoir), we have the map
\begin{align}
{{\hat\varrho}}_{n+1}={\rm Tr}_{\rm HC}
\left\{ \mathcal{\hat P} \mathcal{\hat R}
\left(
{\hat\varrho}_{n}\otimes {\hat\chi} \otimes {\hat\theta}
 \right) 
\mathcal{\hat R}^\dag \mathcal{\hat P}^\dag \right\},
\end{align}
where the trace is over both [hot (H) and cold (C)] reservoirs.


\subsection{Statistical distribution of the battery ergotropy}

We numerically investigate the mean and the statistical distribution of
ergotropy, as a function of the number of collisions with the reservoirs.
Hereafter we set the state of the cold reservoir qubits as 
$\hat{\theta}=|\!\downarrow\, \rangle\langle \,\downarrow \! |$. 
This ideal case of a zero-temperature reservoir leads, after a collision with complete swap, 
to the passive state with the lowest energy (i.e., ground state energy) for the system. 

\begin{figure}[h]
	\includegraphics[width=9.cm]{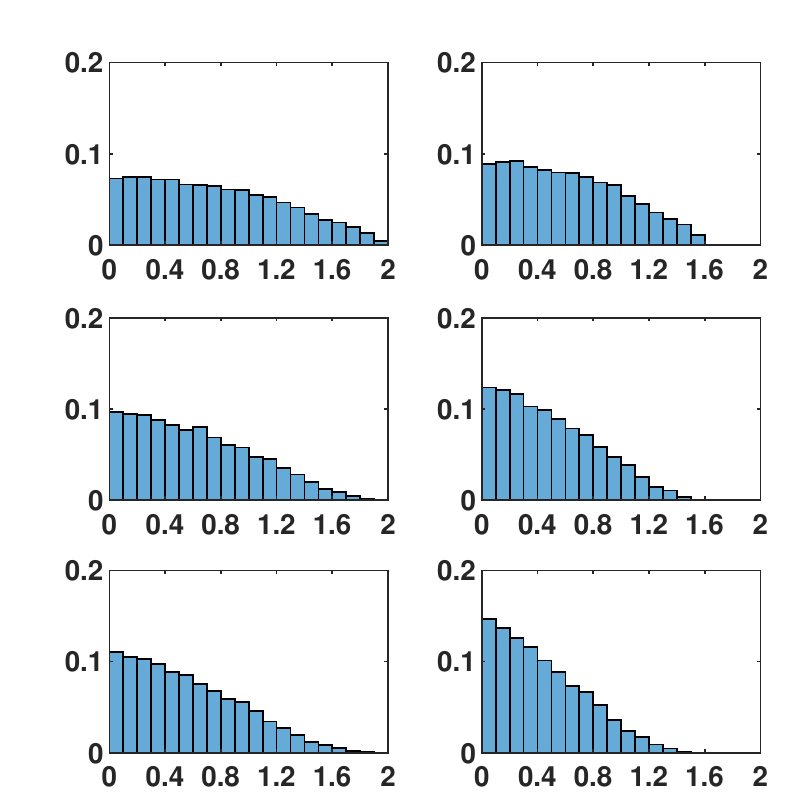}
	\caption{Histograms showing the statistical distribution of the ergotropies of the system, after collisions
	with the hot reservoir (left panels) or the cold reservoir (right panels). From top to bottom:
	distributions after the first, the second, and the third collision with the (hot or cold) reservoir.
	The hot reservoir consists of qubits ($\mu=2$); the swap parameter $\alpha=\frac{\pi}{10}$.}
	\label{fig:dist1}
\end{figure}

We first consider the case in which the hot reservoir consists of qubits
($\mu=2$). Initially the system is prepared in a pure state
(which one is irrelevant when averaging over random collisions). 
In Fig.~\ref{fig:dist1} we show the numerically generated histograms of statistical distributions of the system 
ergotropies, after each of the first three collisions with the hot and the cold reservoir. Hereafter, we consider $10^4$ 
trajectories, i.e., each one with random unitaries drawn 
from the invariant Haar measure on the unitary group $U(L=2\mu)=4$ .

\begin{figure}[h]
	\includegraphics[width=7.cm]{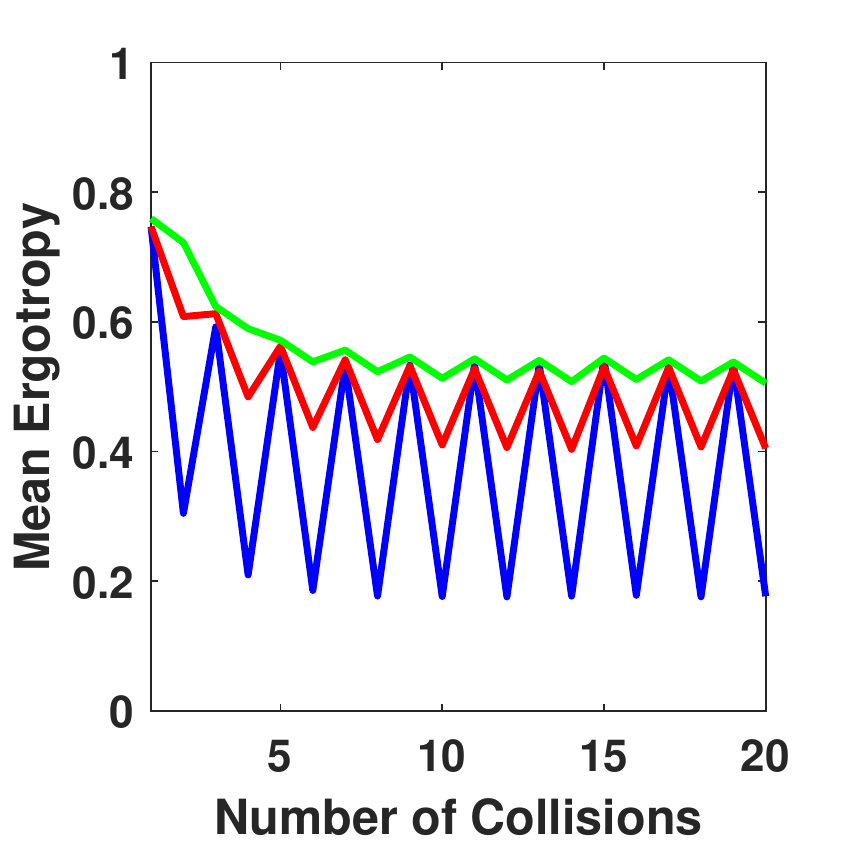}
	\caption{Ensemble-averaged ergotropy as a function of the number of collisions (alternating collisions
	with the hot  -$\mu=2$- and the cold reservoir).  Swap parameter $\alpha=\frac{\pi}{20}$ (green line),
	$\frac{\pi}{10}$ (red line), and $\frac{\pi}{5}$ (blue line).}
	\label{fig:ergoswap}
\end{figure}

We then show in Fig.~\ref{fig:ergoswap} the mean ergotropy as a function of the number of collisions, 
for different values of the swap parameter $\alpha$. We can see that a periodic steady state is approached, 
with a period of two collisions, one with the hot and one with the cold reservoir. 
The value of $\alpha$ affects the time needed to practically achieve the periodic steady state, as well 
as the 
working of the quantum battery. Indeed, the effectiveness of the discharging process 
increases with the swap parameter, and a passive state is obtained for the limiting case of complete swap, $\alpha=\frac{\pi}{2}$.

\begin{figure}[h]
	\includegraphics[width=8.5cm]{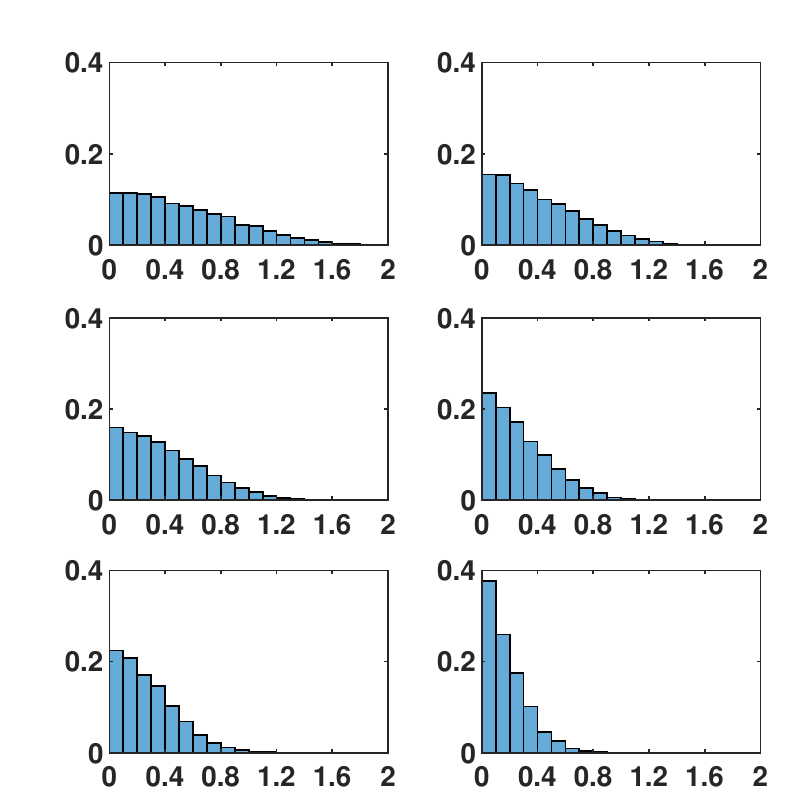}
	\caption{Histograms showing the statistical distribution of the ergotropies of the system, after 6 cycles, a collision 
	with the hot reservoir (left panels) and a further collision with the cold reservoir (right panels). From top to bottom:
	qudits of dimension $\mu=2$, $4$, and $8$. Swap parameter $\alpha=\frac{\pi}{10}$.}
	\label{fig:dist2}
\end{figure}

\begin{figure}[h]
	\includegraphics[width=7.cm]{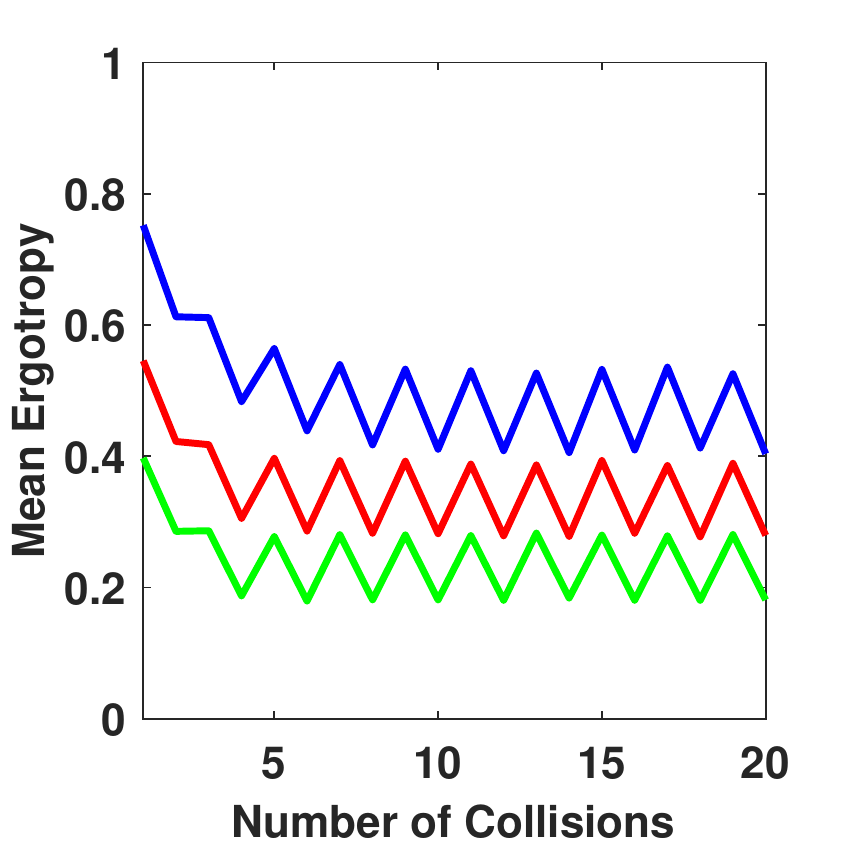}
	\caption{Ensemble-averaged ergotropy as a function of the number of collisions (alternating collisions
	with the hot and the cold reservoir).  Swap parameter $\alpha=\frac{\pi}{10}$, qudits of dimension 
	$\mu=2$ (blue line), $4$ (red line), and $8$ (green line).}
	\label{fig:ergomu}
\end{figure}

In what follows, we change the dimension $\mu$ of the qudits in the hot reservoir. The statistical distributions
of the ergotropies is shown in Fig.~\ref{fig:dist2} after six cycles, so that for the used value of the swap parameter
($\alpha=\frac{\pi}{10}$) the periodic steady-state is in practice achieved.  We can see that the distribution shrinks 
with increasing the dimension $\mu$.  
The mean ergotropy, shown in Fig.~\ref{fig:ergomu} as a function of the number of collisions,
is smaller at larger $\mu$.
Indeed, in the limit $\mu\to\infty$ the non-equilibrium reservoir becomes an
infinite-temperature reservoir, leading the system qubit after each collision to the completely mixed 
thermal state $\varrho=\frac{I}{2}$, for which the ergotropy vanishes.

\section{A quantum Otto engine fuelled by random collisions}

In the previous section we have introduced a model where the hot reservoir is a non-equilibrium reservoir, 
which can enhance coherences in the working medium, which in our case is a qubit system. 
It is therefore appealing to investigate whether these coherences
could be used to improve the performance of a heat engine.
For that purpose, we consider a model of a quantum Otto cycle, consisting of four strokes.

\textit{Stroke A}: the system, initially in the state $\hat{\varrho}$, 
interacts with the hot, non-equilibrium  reservoir, while its Hamiltonian remains unchanged, 
${{\hat{H}_1}}=\frac{1}{2}\Delta_1{\hat{\sigma}}_z$ ($\Delta_1>0$).
The collision is modeled as a random unitary transformation, after which 
the system density matrix becomes $\hat{\varrho}'$.
The heat absorbed by the system is then given by
\begin{align}
Q_{\rm in}={\rm Tr}(\hat{H}_1 \hat{\varrho}')-
{\rm Tr}(\hat{H}_1\hat{\varrho})=\frac{1}{2}\Delta_1(z'-z),
\end{align}
where $z$ and $z'$ are the $z$-components of the Bloch vectors of 
$\hat{\varrho}$ and of $\hat{\varrho}'$, respectively.

\textit{Stroke B}: the system is decoupled from the hot reservoir and its
Hamiltonian is adiabatically changed up to 
$\hat{H}_2=\frac{1}{2}\Delta_2\hat{\sigma}_z$ (with $\Delta_1>\Delta_2>0$), 
whereas the system density matrix $\hat{\varrho}'$ remains unchanged. 
The work performed by the system is given by
\begin{align}
W_{\rm out}={\rm Tr}(\hat{H}_1\hat{\varrho}')-
{\rm Tr}(\hat{H}_2 \hat{\varrho}')= \frac{1}{2}z'(\Delta_1-\Delta_2).
\end{align}

\textit{Stroke C}: the system Hamiltonian remains unchanged, and the 
interaction of the system with the cold, thermal reservoir is modeled by a partial swap collision, 
after which the sytem density matrix becomes $\hat{\varrho}''$.
The heat absorbed by the system (on average negative) is
\begin{align}
Q_{\rm out}={\rm Tr}(\hat{H}_2\hat{\varrho}'')-
{\rm Tr}(\hat{H}_2\hat{\varrho}')=\frac{1}{2}\Delta_2(z''-z').
\end{align}

\textit{Stroke D}: the system is decoupled from the cold reservoir and its Hamiltonian adiabatically
returns to $\hat{H}_1=\frac{1}{2}\Delta_1\hat{\sigma}_z$, whereas the system density matrix  $\hat{\varrho}''$ remain unchanged,. 
The work performed by the system (on average negative) is
\begin{align}
W_{\rm in}={\rm Tr}(\hat{H}_2\hat{\varrho}'')-
{\rm Tr}(\hat{H}_1\hat{\varrho}'')= \frac{1}{2}z''(\Delta_2-\Delta_1).
\end{align}

We point out that each single realization of the Otto cycle is not strictly speaking a true cycle 
since in general the final state $\hat{\varrho}''$  is different from the initial state $\hat{\varrho}$. However, by a suitable ensemble average it is  possible to define a ``macroscopic'' efficiency. 

If we use the standard formula  for the efficiency $\eta$ of a heat engine we obtain
\begin{align}
\eta=\frac{W}{Q_{\rm in}} =
\frac{z'-z''}{z'-z}\left(1-\frac{\Delta_2}{\Delta_1}\right),
\end{align}
where $W=W_{\rm in}+W_{\rm out}$.
After ensemble averaging, we obtain a periodic steady-state (with the period of the Otto cycle), 
and therefore $\langle \hat{\varrho}''\rangle=\langle 
\hat{\varrho}\rangle$ ($\langle\dots  \rangle$ denotes ensemble averaging).
Consequently we have $\langle z''\rangle =\langle z \rangle$, which leads to the 
standard Otto cycle efficiency:
\begin{align}
\eta_m=\frac{\langle W \rangle}{\langle Q_{\rm in}\rangle}=1-\frac{\Delta_2}{\Delta_1}.
\end{align}
This macroscopic efficiency can be 
obtained after averaging work and input heat over a large number of cycles
and/or over an ensemble of random collisions.
While efficiency assumes a clear thermodynamic meaning only after one of these two averages, it is nevertheless
interesting, when considering the engine constancy, to investigate efficiency fluctuations.
In what follows, we shall perform such study.

\begin{figure}[h]
	\includegraphics[width=7.5cm]{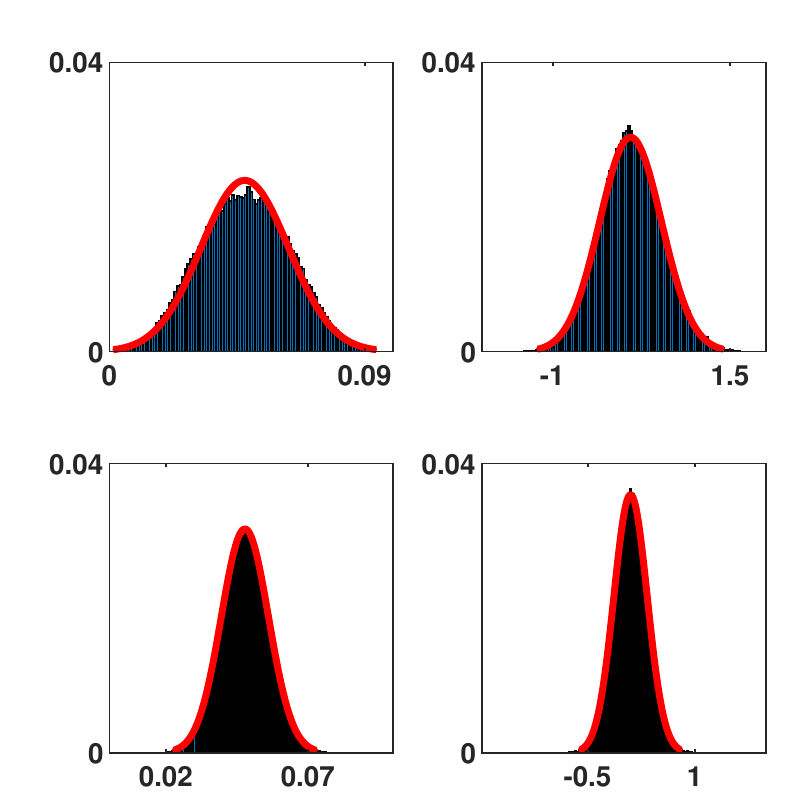}
	\caption{Histogram showing the statistical distribution of  work $W$ (left panels) and input heat $Q_{\rm in}$ (right panels)
	for the Otto cycle described in the text.
	The hot reservoir consists of qudits of dimension $\mu=2$ (top panels) and $8$ (bottom panels).
	Swap parameter $\alpha=\frac{\pi}{10}$, qubit gaps during the cycle: $\Delta_1=2$, $\Delta_1=1$. 
	Red lines show Gaussian fits. }
	\label{fig:workheat}
\end{figure}

We first consider the statistical distributions of work $W$ and input heat $Q_{\rm in}$ , shown in Fig.~\ref{fig:workheat}.
Hereafter, histograms are constructed on $10^5$ cycles.
It can be seen that all histrograms are nicely fitted by a Gaussian distribution, 
of width decreasing with increasing the dimension $\mu$ of the qudits
in the non-equilibrium reservoirs.  
More precisely, we have seen that the standard deviations of both $W$ and $Q_{\rm in}$ 
decrease, for a given swap parameter $\alpha$, as $1/\sqrt{\mu}$.
We have also checked the validity of the Gaussian fit for other values of $\alpha$, 
with the standard deviation of the work distribution increasing with $\alpha$.

\begin{figure}[h]
	\includegraphics[width=7.5cm]{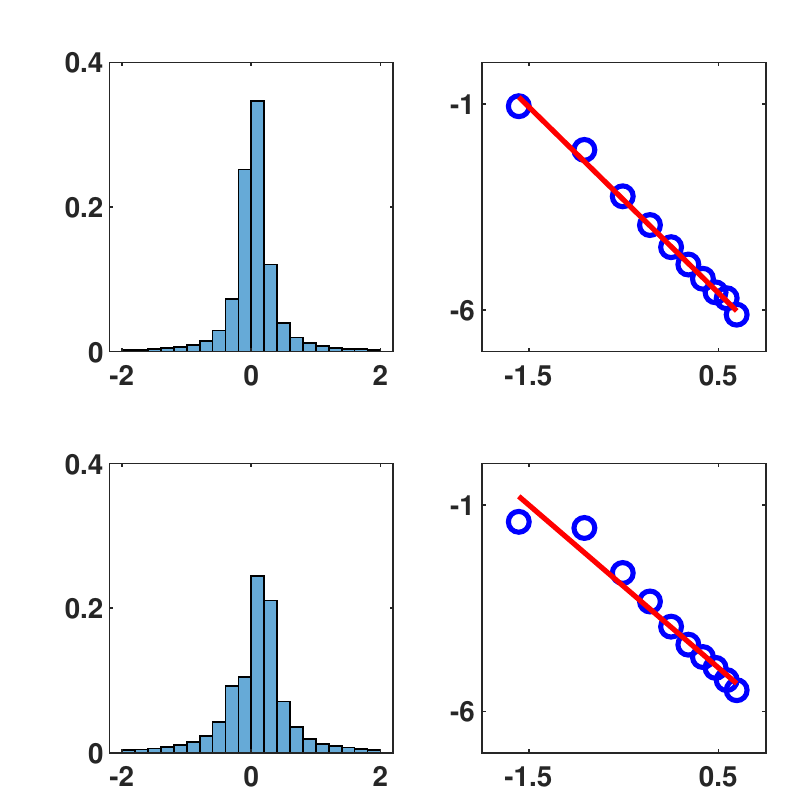}
	\caption{Histogram showing the statistical distribution of  the efficiency $\eta$,
	for dimension of the hot reservoir qudits $\mu=2$ (top panels) and $8$ (bottom panels),
	for the same parameter  values as in Fig.~\ref{fig:workheat}.
	Right panels show the $\eta>0$ data in log-log scale (natural logarithms), with power law fits 
	$p(\eta)\propto 1/\eta^\alpha$, with $\alpha=-2.25$ (top) and $-1.97$ (bottom).} 
	\label{fig:efficiency}
\end{figure}

Finally, we consider the efficiency distributions, shown (for the cases of 
Fig.~\ref{fig:workheat}) in Fig.~\ref{fig:efficiency}. Assuming, as confirmed by numerical simulations, Gaussian distributions 
for both $W$ and $Q_{\rm in}$, we obtain a ratio distribution $p(\eta)$ for $\eta=W/Q_{\rm in}$ known from the literature 
(see appendix for details).
Such distribution predicts $p(\eta\rightarrow\pm\infty)\propto\eta^{-2}$, a decay consistent with the power-law fits
shown in the right plots of Fig.~\ref{fig:efficiency}.
Since the standard deviation of both $W$ and $Q_{\rm in}$ drops with $\mu$ as  $1/\sqrt{\mu}$, we can conclude that
the statistical distribution of the efficiency is independent of the size of qudits in the non-equilibrium reservoir. 
Finally, we point out that a decay $p(\eta\rightarrow\pm\infty)\propto\eta^{-2}$ implies that moments of any order of
$p(\eta)$ are not finite. In particular, the mean efficiency $\langle \eta\rangle$ is not finite and therefore does not 
coincide with the macroscopic efficiency $\eta_m=\langle W \rangle/\langle Q_{\rm in}\rangle$~\cite{Esposito2015}.

\section{Conclusions}

In this paper, we have characterized the working of the smallest size quantum medium, i.e. a single qubit, 
when operating either as a quantum battery or as a working fluid in a quantum Otto cycle. 
While we have considered a standard, thermal reservoir as a cold bath, we have modeled the action of the hot, 
non-equilibrium reservoir via random collision. 
This setup allows to extract work, with 
the mean, macroscopic efficiency equal to the one obtained with
the hot reservoir being a thermal one. 
On the other hand, fluctuations due to the random nature of interactions 
play a very important role. 
In particular, the distribution of efficiencies does not afford finite moments of any order, 
so that the mean of efficiencies does not 
coincide with the macroscopic efficiency. 

It is interesting to remark that random collisions, that is, operations which by definition do not require specific control 
of the unitary transformations, can be used to extract work. It would be interesting to explore such possibility in 
real, noisy intermediate-scale quantum hardware or in quantum annealers.    

\acknowledgements
We acknowledge support by the INFN through the project QUANTUM.
Vahid Shaghaghi acknowledges the receipt of a fellowship from the ICTP Programme for Training and Research in Italian Laboratories, 
Trieste, Italy.

\appendix

\begin{widetext}

\section{Efficiency probability distribution}
Assuming that the work $W=W_{\rm in}+W_{\rm out}$ and the input heat $Q_{\rm in}$ are independent Gaussian random variables, 
we can obtain an analytical expression the efficiency p.d.f. $p(\eta)$. Following~\cite{Curtiss1941,Hinkley1969}, we have
\begin{align}
p(\eta)=\int d W dQ \,\delta\left(\eta-\frac{W}{Q}\right) p(W,Q)
= \int_{-\infty}^{\infty}\left|Q\right| p(\eta Q,Q) dQ,
\end{align}
where $p(W,Q)$ is the joint probability distribution function for $W$ and $Q_{\rm in} $ (to simplify writing, hereafter we set 
$Q\equiv Q_{\rm in}$):
\begin{align}
p(W,Q)=\frac{1}{2\pi\sigma_Q\sigma_w}\exp\left\{-\frac{1}{2}\left[\frac{\left(W-\mu_w\right)^2}{\sigma_W^2}+\frac{\left(Q-\mu_Q\right)^2}{\sigma_Q^2}\right]\right\},
\end{align}
with $\mu_W$, $\mu_Q$ and $\sigma_W$, $\sigma_Q$ mean and standard deviation of the Gaussian distributions
for $W$ and $Q$.
After straightforward integration we obtain
\begin{align}
p(\eta)=\frac{d(\eta)b(\eta)}{2\sqrt{2\pi}\sigma_Q\sigma_w{a(\eta)}^3}\left[2-{\rm erf}\left(-\frac{b(\eta)}{\sqrt{2}a(\eta)}\right)+{\rm erf}\left(\frac{b(\eta)}{\sqrt{2}a(\eta)}\right)\right]+\ \frac{e^{-\frac{c}{2}}}{\pi\sigma_Q\sigma_w{a(\eta)}^2},
\end{align}
where we have introduced
\begin{align}
a(\eta)=\sqrt{\frac{\eta^2}{\sigma_w^2}+\frac{1}{\sigma_Q^2}},
\quad
b(\eta)=\frac{\mu_w\eta}{\sigma_w^2}+\frac{\mu_Q}{\sigma_Q^2},
\quad
c=\frac{\mu_w^2}{\sigma_w^2}+\frac{\mu_Q^2}{\sigma_Q^2} ,
\quad
d(\eta)=exp\left\{\frac{b^2(\eta)-ca^2(\eta)}{2a^2(\eta)}\right\}.
\end{align}
This distribution has power-law tails: $p(\eta\rightarrow\pm\infty)\propto\eta^{-2}$.
\end{widetext}


\end{document}